\documentclass[12pt]{article}
\usepackage{amsmath,epsfig,graphicx,amssymb}

\newcommand{\be}{\begin{equation}}
\newcommand{\ee}{\end{equation}}
\newcommand{\ben}{\begin{eqnarray}}
\newcommand{\een}{\end{eqnarray}}
\newcommand{\bes}{\begin{subequations}}
\newcommand{\ees}{\end{subequations}}
\newcommand{\bF}{\begin{figure}}
\newcommand{\eF}{\end{figure}}

\begin{document}

\title{The Central Mystery of Quantum Mechanics}
\author{Partha Ghose\footnote{partha.ghose@gmail.com} \\
Centre for Natural Sciences and Philosophy, \\ 1/AF Bidhan Nagar,
Kolkata, 700 064, India\\ and\\Centre for Philosophy and Foundations of Science,\\Darshan Sadan, E-36 Panchshila Park,
New Delhi 110017, India}

\maketitle

\begin{abstract}
A critical re-examination of the double-slit experiment and its variants is presented to clarify the nature of what Feynmann called the ``central mystery'' and the ``only mystery'' of quantum mechanics, leading to an interpretation of complementarity in which a `wave {\em and} particle' description rather than a `wave {\em or} particle' description is valid for the {\em same} experimental set up, with the wave culminating in the particle sequentially in time. This interpretation is different from Bohr's but is consistent with the von Neumann formulation as well as some more recent interpretations of quantum mechanics.
\end{abstract}

\section{INTRODUCTION}
Thomas Young is believed to have carried out his famous double-slit experiment with light in 1801. It was one of the key experiments that helped overthrow the corpuscular theory of light favoured by Newton and establish the wave theory favoured by Christiaan Huygens and Augustin Fresnel. However, when classical light was replaced by single-photon/particle sources and detectors in the 20th century, the old wave-particle controversy was revived because it revealed a mysterious dual character of light. According to Feynman it contains ``the heart of quantum mechanics. In reality, it contains the only mystery. We cannot make the mystery go away by explaining how it works . . . '' \cite{feynman}. Much has been written on this mystery both technically, in text books and in popular literature. The statements made by the founding fathers of quantum mechanics like Einstein, Bohr, Heisenberg and von Neumann who held different positions, heighten the mystery. 

The purpose of this paper is to bring together some of these conflicting points of view not usually found in pedagogical texts, together with some comments made by later reputed thinkers like Richard Feynman, John Bell and Roger Penrose, in order to arrive at an informed and consistent understanding of the mystery that still persists in spite of much later work such as {\em welcher-weg} experiments \cite{ww}, delayed-choice experiments \cite{wheeler} and theoretical interpretations involving hidden variable theories \cite{bohm}, path-integrals \cite{feynman2}, many worlds \cite{everett}, consistent histories \cite{griffiths} and decoherence \cite{zurek}. No attempt is made, however, to cover all these important areas of later work in foundations of quantum mechanics including quantum information processing. The emphasis is rather on making clear what the central and only mystery of standard quantum mechanics really is, using the familiar double-slit experiment and its variants, and thus motivating these later developments.  

The plan of the paper is as follows. Section 2 introduces the double-slit experiment and discusses Bohr's original interpretation of wave-particle duality in terms of his Complementarity Principle and also mentions Heisenberg's interpretation in terms of the uncertainty principle. Section 3 deals with von Neumann's formulation of quantum mechanics which is different from the Bohr-Heisenberg formulation and which led to the notorious `measurement problem'. In section 4 the double-slit experiment is revisited together with modern variants of it such as {\em welcher-weg} experiments, delayed-choice experiments and quantum erasures to prepare the ground for Section 5 in which an interpretation of complementarity that is different from Bohr's but consistent with von Neumann's formulation as well as some more recent interpretations of quantum mechanics is developed and spelt out without, of course, solving the mystery which persists in spite of many claims and counterclaims.

\section{THE DOUBLE-SLIT EXPERIMENT: COMPLEMENTARITY AND UNCERTAINTY}

An example of sound popular exposition of the double-slit experiment based on interpretations by some of the founding fathers of quantum mechanics such as Bohr is the following quotation \cite{gribbin}: 
\begin{quote}

When a beam of monochromatic light is shone through two narrow holes in a screen, the light spreading out from the two holes interferes, just like ripples interfering on the surface of a pond, to produce a characteristic pattern on a second screen. 

The mystery is that light can also be described as a stream of particles, called photons. The light source in a Young's slit experiment can be turned down to the point where it consists of individual photons going through the experiment, one after the other. If the spots of light made by individual photons arriving at the second screen (actually a photoelectric detector) are added together, they still form an interference pattern, as if each photon goes through both holes and interferes with itself on the way through the experiment. 
\end{quote}
Any attempt to determine which hole the photon goes through, however, destroys the interference pattern. A formal way of seeing this without invoking the uncertainty principle is to write the normalized state of the photon after passage through the double-slit as a coherent superposition of the two normalized states generated by the two slits (Fig. 1),

\be
\vert \psi\rangle = \frac{1}{\sqrt 2} [\vert \psi\rangle_{\rm{S_1}} + e^{i \theta}\vert \psi\rangle_{\rm{S_2}}].
\ee
where $\theta$ is the phase difference between the two states. The interference between the states is then given by
\be
\langle \psi\vert \psi\rangle = 1 + \, _{\rm{S_2}}\langle \psi\vert \psi\rangle_{\rm{S_1}} \rm{cos} \theta.
\ee
The variation of the phase difference $\theta$ along the second screen produces an interference pattern on it characteristic of wave propagation. If detectors are placed to observe which path the photon definitely takes, then the state of the
photon {\em plus} the detectors becomes

\begin{figure}[ht]
\begin{picture}(100,100)(-100,0) 
\put(1,1){\line(0,1){90}}
\put(0,-100){\line(0,1){90}}
\put(143,-100){\line(0,1){200}}
\put(0,-5){\line(1,1){60}}
\put(0,-4){\line(1,-1){60}}
\put(80,57){\line(1,-1){55}}
\put(80,-65){\line(1,1){55}}
\put(70,-60){\line(0,1){110}}
\put(70,65){\line(0,1){34}}
\put(70,-100){\line(0,1){34}}
\put(-35,-7){$\rm{Source}$}
\put(70,-110){$S$}
\put(143,-3){\circle*{4}}
\put(65,55){$S_1$}\put(65,-70){$S_2$}
\put(115,-110){$\rm{Second\, Screen}$}
\put(6,6){\vector(1,1){15}}
\put(6,-15){\vector(1,-1){15}}
\put(80,-70){\vector(1,1){15}}
\put(80,63){\vector(1,-1){15}}
\put(-90,-140){Fig. 1 The double-slit set up with two slits $S_1$ and $S_2$ in a screen $S$}
\end{picture}\vspace{2in}\end{figure}

\be
\vert \phi\rangle = \frac{1}{\sqrt 2} [\vert \phi\rangle_{\rm{S_1}}\vert D_1\rangle + e^{i \theta}\vert \phi\rangle_{\rm{S_2}}\vert D_2\rangle].
\ee
Each of the detectors $D_i$ ($i = 1,2$) fires $50$ per cent of the time and they always do so in anti-coincidence. In each of these cases it is usually claimed that it is possible to trace the path of the photon from the slit whose associated detector fired back to the source. In all these cases the photon certainly did not pass through the other slit. Since the state of the detector that fired is orthogonal to that of the detector that did not fire,
\be
\langle \phi\vert \phi\rangle = 1
\ee
and the interference term disappears. This is usually taken to imply that there is no observable evidence of wave-like behaviour in these cases. Bohr \cite{bohr} writes,
\begin{quote}
This point is of great logical consequence, since it is only the circumstance that we are presented with a choice of {\em either} tracing the path of a particle {\em or} observing interference effects, which allows us to escape from the paradoxical necessity of concluding that the {\bf behaviour} of an electron or a photon should depend on the presence of a slit in the diaphragm through which it could be proved not to pass. We have here to do with a typical example of how the complementary phenomena appear under mutually exclusive experimental arrangements (cf. p. 210) and are just faced with the impossibility, in the analysis of quantum effects, of drawing any {\bf sharp separation between an independent behaviour of atomic objects and their interaction with the measuring instruments} which serve to define the conditions under which the phenomena occur. (bold face type added)
\end{quote}
This is the essence of Bohr's Complementarity Principle as applied to the double-slit arrangement in which one observes {\em either} a double-slit interference pattern on the second screen characteristic of wave-like propagation when both the slits are open, {\em or} the absence of it when detectors are placed near the slits, which is taken to be characteristic of particle-like behaviour. The {\em description} of the photon thus depends on whether both the slits are kept open so that {\em in principle} no path information is available, or detectors are placed near the slits so that {\em in principle} `which-path' information is available. This is the basis of `wave-particle duality' -- in a double-slit arrangement one can consistently {\em describe} an atomic object {\em either} as a wave {\em or} as a particle but never both simultaneously in the same experimental set up. There is no ontological duality in the object itself, its {\em behaviour} being determined by its `inseparable' interaction with the measuring instrument. In Bohr's words \cite{bohr2},
\begin{quote}
{\em Complementarity}: any given application of classical concepts precludes the simultaneous use of other classical concepts which in a different connection are equally necessary for the elucidation of the phenomena.
\end{quote} 
Another important feature of the measuring apparatus, according to Bohr, is its classical nature. According to him \cite{bohr3},
\begin{quote}
The experimental conditions can be varied in many ways, but the point is that in each case we must be able to communicate to others what we have done and what we have learned, and that therefore {\em the functioning} of the measuring instruments must be described within the framework of {\em classical physical ideas}. (italics added)
\end{quote}
Thus, not only is it impossible to draw any sharp separation between independent atomic objects and the measuring instruments, the latter must {\em function} classically. Since, according to Bohr, the description of observations and communication with others must both be in the language of classical physics, one might ask: where is the room for non-classical behaviour? Bohr's answer is: it lies in the Complementarity Principle. Armed with such a principle, Bohr is believed to have been able to interpret wave-particle duality, not as an ontological duality which would be self-contradictory in classical/ordinary language, but as a {\em duality of behaviours} under mutually exclusive experimental conditions. Hence, Bohr's use of the word `complementarity' is not to be confused with complementarity in the usual sense of the word. Einstein had problems in understanding Bohr's formulation of this principle. He wrote \cite{schilpp}, 
\begin{quote}
Despite much effort which I have expended on it, I have been unable to achieve the sharp formulation of Bohr's principle of complementarity.
\end{quote}
On the other hand, John Bell who, like Einstein, was known for his strong advocacy of a realist interpretation of quantum mechanics, offered the following sympathetic and interesting interpretation of Bohr's nonrealist point of view \cite{bell1} :
\begin{quote}
It seems to me that Bohr used this word with the reverse of its usual meaning. Consider for example the elephant. From the front she is head, trunk and two legs. From the back she is bottom, tail, and two legs. From the sides she is otherwise, and from the top and bottom different again. These various views are complementary in the usual sense of the word. They supplement one another, they are consistent with one another, and they are all entailed by the unifying concept `elephant'. It is my impression that to suppose Bohr used the word `complementary' in this ordinary way would have been regarded by him as missing his point and trivializing his thought. He seems to insist rather that we must use in our analysis elements which {\em contradict} one another, which do not add up to, or derive from, a whole. By `complementarity' he meant, it seems to me, the reverse: contradictoriness. Bohr seemed to like aphorisms such as `the opposite of a deep truth is also a deep truth'; `truth and clarity are complementary'. Perhaps he took a subtle satisfaction in the use of a familiar word with the reverse of its meaning.

`Complementarity' is one of what might be called the `romantic' world views inspired by quantum theory. It emphasizes the bizarre nature of the quantum world, the inadequacy of everyday notions and classical concepts. It lays stress on how far we have left behind naive 19th century materialism. 
\end{quote}
This was regarded by some as the new `conceptual enlightenment' brought about by Bohr of the abstract mathematical structure of quantum mechanics developed by the G\"{o}ttingen school led by Born, Jordan and Heisenberg with Schr\"{o}dinger, Dirac and Pauli joining in. In spite of all this impressive mathematical innovation none had however succeeded to describe so simple a phenomenon as the path of an electron in the cloud chamber. 

Heisenberg, however, sought the answer to the problem in a different direction from Bohr, namely in terms of his uncertainty relation $\Delta x \,\Delta p \geq \hbar/2$. This is how he describes the situation \cite{heisenberg}
\begin{quote}
I had meanwhile been so far educated by the G\"{o}ttingen mathematical school as to assume that, through logical application of the quantum-mechanical formalism, conclusions must also be inferrable as to he remainder of the old concepts that would survive in the new language. But Bohr wanted to set out from the two initially contradictory pictures of the wave and the corpuscular theories, and to push on from thence to the correct concepts. The answer was then, as you know, made possible by reversing the statement of the problem; the question was no longer to be, ``How do we represent the path of the electron in the cloud chamber?''; instead, we had to ask, ``Are there, perhaps, in the observation of nature, only such experimental situations as can be represented in the mathematical formalism of quantum theory?'' Is it, in other words, correct, as Einstein once maintained against me, that theory first decides what can be observed? The answer could then be given in the form of the uncertainty-relation. The concept of path may be used only with the degree of inexactness characterised by the fact that the product of the uncertainty of position and the uncertainty of the associated momentum cannot be smaller than Planck's quantum of action.
Bohr had arrived at the same limitations of language by way of the concept of complementarity formulated by him, and only now was it possible to state clearly what we are to understand by an observation-situation, and how it is represented in the mathematical formalism. Pauli at once concurred in this interpretation.
\end{quote}
 
\section{THE VON NEUMANN FORMULATION}

An unsatisfactory feature of the Bohr interpretation is its insistence that `the very small and the very big must be described in very different ways, in quantum and classical terms respectively' \cite{bell2}. This in spite of the fact that there is no intrinsic scale within quantum theory to distinguish the very small from the very big. It also requires the old classical physics even to {\em define} the new theory which is supposed to replace it. Furthermore, a classical apparatus is supposed to be made up of atomic constituents which must themselves behave quantum mechanically to be consistent. How can then an apparatus behave classically? As John Bell put it, the `apparatus should not be separated off from the rest of the world into black boxes, as if it were not made of atoms and not ruled by quantum mechanics' \cite{bell3}. 

John von Neumann had realized all this much earlier and sought to alter this situation by introducing a different `romantic' world view into physics in which the essential distinction is not between the very small and the very big but between `matter' and `mind' \cite{neumann}.$^1$ He did this by bringing {\em all} physical systems, the very small and the very big, including the measuring apparatus, within the realm of quantum mechanics. 
With this world view in mind von Neumann introduced two different processes, the usual Schr\"{o}dinger process of evolution of the $\psi$-function which is unitary and reversible and universally applicable to {\em all} things material and physical, small {\em and} big (which he called `process 2'), and a new process which he called `process 1' (also called `collapse' or `reduction' of the $\psi$-function) which is essentially {\em non-quantum mechanical}, non-unitary, irreversible, and {\em extra-physical and perceptual} and {\em not reducible to the physical environment}. This is an aspect of von Neumann's `process 1' that is often overlooked in the literature, and therefore needs to be emphasized. Another important technical point to emphasize about `process 1' is that it changes a non-diagonal density matrix of a pure state into a `reduced' density matrix of a mixed state that is diagonal and whose diagonal terms can be interpreted as probabilities. This is achieved by means of the `projection postulate' which is the mathematical expression for `process 1'. To quote von Neumann \cite{neumann}: 
\begin{quote}
The difference between these two processes $U \rightarrow U^\prime$ is a very fundamental one: aside from the different behaviors in regard to the principle of causality, they are also different in that the former is (thermodynamically) reversible, while the latter is not (cf. V.3).

Let us now compare these circumstances with those which actually exist in nature or in its observation. First, it is inherently entirely correct that the measurement or the related process of the subjective perception$^2$ is a new entity relative to the physical environment and is not reducible to the latter. Indeed, subjective perception leads us into the intellectual inner life of the individual, which is extra-observational by its very nature (since it must be taken for granted by any conceivable observation or experiment). $\dots$ Nevertheless, it is a fundamental requirement of the scientific viewpoint - the so called principle of the psycho-physical parallelism - that it must be possible to describe the {\em extra-physical process of the subjective perception} as if it were in reality in the physical world - i.e., to assign to its parts equivalent physical processes in the objective environment, in ordinary space. $\cdots$ But in any case, no matter how far we calculate -- to the mercury vessel, to the scale of the thermometer, to the retina, or into the brain, at sometime we must say: and this is perceived by the observer. That is, we must always divide the world into two parts, the one being the observed system, the other being the observer. In the former, we can follow up all physical processes (in principle at least) arbitrarily precisely. {\em In the latter, this is meaningless}. The boundary between the two is arbitrary to a very large extent. … That this boundary can be pushed arbitrarily deeply into the interior of the body of the actual observer is the content of the principle of psycho-physical parallelism - but this does not change the fact that in each method of description the boundary must be put somewhere, if this method is not to proceed vacuously, i.e., if a comparison with experiment is possible. {\em Indeed, experience only makes statements of this type: an observer has made a certain (subjective) observation; and never any like this: a physical quantity has a certain value}. $\cdots$

Now quantum mechanics describes the events which occur in the observed portions of the world, so long as they do not interact with the observing portion, with the aid of process 2. (V.1), but as soon as such an interaction occurs, i.e., a measurement, it requires the application of process 1. {\em The dual form is therefore justified}. However, the danger lies in the fact that the principle of the psycho-physical parallelism is violated, so long as it is not shown that the boundary between the observed system and the observer can be displaced arbitrarily … (pp. 418-421; emphases added) 
\end{quote} 
Unfortunately, this is a formulation in which two distinctly different processes not reducible one to the other, contribute to the overall situation in {\em physics}, with the boundary between them inherently `shifty'. Roger Penrose calls these processes ${\bf U}$ and ${\bf R}$, ${\bf U}$ standing for unitary evolution and ${\bf R}$ for reduction of the state vector \cite{penrose}. This inherent ambiguity in the very theoretical foundation of quantum mechanics resulted in the notorious `measurement problem' or `measurement paradox', a point forcefully made by John Bell according to whom \cite{bell5},
\begin{quote}
The first charge against `measurement', in the fundamental axioms of quantum mechanics, is that it anchors there the shifty split of the world into `system' and `apparatus'. A second charge is that the world comes loaded with meaning from everyday life, meaning which is entirely inappropriate in the quantum context. 
\end{quote} 
In whichever way you look at it, therefore, in Bohr's way or in von Neumann's way or in any of the other ways found in the literature (Dirac, Landau and Lifshitz, Gottfried, van Kampen) and critiqued by Bell \cite{bell6}, the ambiguity remains. 

It has been argued by some that von Neumann's `projection postulate' is not an essential part of quantum mechanics. This view regards the postulate as  an unphysical process and an ``optional discarding of certain branches of the state vector that are expected to be
irrelevant for the purpose at hand'' \cite{ballentine} because all statistical predictions of quantum mechanics concerning correlated multi-particle systems can be derived without using this postulate. A careful reading of von Neumann shows that he, too, did not regard `process 1' as a conventional physical process but rather as an {\em extraphysical process of subjective perception}. But he certainly did not regard it as being optional. A convincing empirical reason why this process cannot, in fact, be optional is the occurrence of {\em individual events} like spots on photographic plates and discrete and countable clicks in detectors that are incompatible with purely unitary evolution of a quantum state. They do occur and eventually produce the statistical patterns predicted by quantum mechanics in the limit of indefinitely large numbers of them. 

Different points of view regarding quantum mechanics often arise because physicists adopt fundamentally different philosophical positions regarding the ontology of the wave function which determines whether one regards the `measurement problem' as a genuine problem of physics or an unnecessary baggage that can be discarded. For example, Ballentine's position follows from his epistemological interpretation of the wave function, namely that ``the state vector is not itself an element of reality, but is only a means to calculate the probability distributions for various observables''. As Roger Penrose writes \cite{penrose}:
\begin{quote}
It is a common view among many of today's physicists that quantum mechanics provides us with {\em no} picture of `reality' at all! The formalism of quantum mechanics, on this view, is to be taken as just that: a mathematical formalism. This formalism, as many quantum physicists would argue, tells us essentially nothing about an actual {\em quantum reality} of the world, but merely allows us to compute probabilities for alternative realities that might occur. Such quantum physicists' ontology -- to the extent that they would be worried by matters of `ontology' at all -- would be the view (a): that there is simply no reality expressed in the quantum formalism. At the other extreme, there are many quantum physicists who take the (seemingly) diametrically opposite view (b): that the unitarily evolving quantum state completely describes actual reality, with the alarming implication that practically all quantum alternatives must always continue to exist (in superposition). $\cdots$ the basic difficulty that confronts quantum physicists, and that drives many of them to such views, is the conflict between the two quantum processes ${\bf U}$ and ${\bf R}$, $\cdots$
\end{quote}
The last phrase is interesting. To von Neumann `process 1' was fundamentally non-quantum mechanical. So, there is an important difference between the way von Neumann viewed his `process 1' and the way Penrose would prefer to look at his process ${\bf R}$.

For other subtle differences between Bohr's interpretation and the `standard' interpretations of quantum mechanics see Gomatam \cite{gomatam}. 

\section{THE DOUBLE-SLIT REVISITED}

With this background let us now have a closer look at the double-slit experiment and its interpretation in terms of `wave-particle duality' and complementarity. With the two slits open an interpretation of the double-slit interference pattern in terms of waves would seem to be the most natural one, and there is total unanimity on this. What about the case in which a detector is placed immediately after one of the slits, say $S_1$? In this case, the path of the particle that passes through $S_1$ can be determined with certainty in $50$ per cent of the cases and the double-slit interference pattern on the second screen disappears. But what about the pattern created by the other slit that is unobstructed? How does it look? And what does it tell us? Most text books are either silent or ambiguous about this. Let us see, for example, how Richard Feynman describes the situation in his famous {\em Lectures on Physics}, Volume 3. He devotes pages 1398-1431 to a long and detailed account of the double-slit experiment in various cases illustrated by several figures. The figures 1-1(b) for bullets, 1-2(b) for water waves, 1-3(b) for electrons, 1-4(b) for electrons that are being watched, 3-1(b) again for electrons, all look the same, which is confusing particularly when he makes it clear from the beginning that: ``Electrons always arrive in identical lumps'' (page 1402).
Let us look at his Fig. 3-4 on page 1431 which is for an ``experiment to determine which hole the electron goes through'' and which gives the probabilities for the electron to arrive at various places on the final screen for three cases. He says, 
\begin{quote}
First of all, if $b$ is zero -- which is the way we would like to design the apparatus -- then the answer is $\cdots$ the probability distribution that you would get if there were only one hole -- as shown in the graph of Fig. 3-4(a). 
\end{quote}   
If one compares Fig. 3-4(a) with Fig. 1-1(b) and Fig. 1-5(b) which show the distribution curve for classical bullets passing through a double-slit arrangement, one could infer from their similarity that electrons passing through a single slit behave like classical bullets. 

The fact of the matter, however, is that with only one slit open, what one observes in experiments, and indeed what quantum mechanics predicts, is not a distribution pattern similar to that of bullets but {\em a single-slit diffraction pattern} characteristic of waves, a fact nowhere stated or illustrated by Feynman. There is good reason therefore to emphasize that even with only one very narrow slit unobstructed in a double-slit arrangement one does not observe particle-like behaviour of the entire ensemble of electrons or photons. What one can claim to observe is their particle-like behaviour in $50$ per cent of the cases (in which a particle {\em is} detected at one of the slits) and wave-like behaviour in the other $50$ per cent of the cases. 

Even this, however, is not the full story because, unlike in the case of water waves and classical optics, with atomic objects the double-slit interference pattern and the single-slit diffraction pattern are both built up over time by individual localized spots created on the screen by the objects -- ``Electrons always arrive in identical lumps'', as Feynman emphasizes. In other words, even in these cases what one {\em observes} are only localized spots which are particle-like. The wave characteristic is reflected in the primacy of the wave function, the probability amplitude, which follows `process 2' but is not directly observable! Hence, the Complementarity Principle must be rephrased for a proper interpretation of the actual experimental conditions. But before attempting that, let us look at a few interesting variants of the double-slit experiment.

\subsection{Welcher-weg experiments}

Rauch and his group initiated a number of experiments in neutron interferometry [2a and 2b] that are variations of the double-slit experiment with static (stochastic) and time-dependent (deterministic) absorbers placed in one path of a neutron interferometer. One of the neutron beams (say the left beam) is chopped by a rotating toothed wheel, and every time the beam is totally obstructed by the wheel, the other beam (the right beam) remains unobstructed (and one has therefore knowledge of which path the neutron takes), whereas at all other times both the beams propagate unattenuated and one has no `which path' knowledge. The probability of the event is therefore the sum of the separate probabilities for each alternative. The speed of the chopper is so adjusted that on the average the left beam undergoes the same attenuation as in the static case. The intensities in the two cases are therefore given by
\ben 
{\rm Stochastic}\nonumber\\
I_0 &=& a \vert \psi_L\vert^2 + \vert \psi_R\vert^2 + 2 \sqrt{a} \psi_L \psi_R {\rm cos} \chi\\
{\rm Deterministic}\nonumber\\
I_0 &=& a \vert \psi_L\vert^2 + \vert \psi_R\vert^2 + 2 a \psi_L \psi_R {\rm cos} \chi
\een where $a$ is the absorption coefficient ($a < 1$) and $\chi$ the phase difference between the paths. The observed intensities were in agreement with these quantum mechanical predictions although intuitively and in classical terms it is hard to understand why the intensities should be different in the two cases. What is also remarkable is that even when 99 per cent of the left beam was blocked, the interference pattern persisted with the same visibility, i.e., contrast.

It has been argued that one can generalize the complementarity idea and give it a quantitative form which would allow one to pass continuously from particle-like information to wave-like information by using information theoretic concepts [2c], and that these experiments provide evidence of that [2d]. Let the initial amplitudes of the two beams be $a$ and let the left beam amplitude be $b$ after absorption. Then the neutron intensity at one of the detectors is
\be
I_0 = a^2 + b^2 + 2 ab\, {\rm cos} (2 k_x x + \phi)
\ee where $k_x$ is the momentum along the $x$ direction. One can now define two parameters $W$ and $P$ by
\ben
W &=& \frac{2 ab}{a^2 + b^2} = {\rm sin} 2 \beta\\
P &=& \frac{a^2 - b^2}{a^2 + b^2} = {\rm cos} 2 \beta
\een
with $a = R {\rm cos} \beta$ and $b = R {\rm sin} \beta$ such that

\be
P^2 + W^2 = 1. 
\label{wp}\ee
When $\beta = \pi/4$, $R =\sqrt{2} a = \sqrt{2}b$ and $W = 1$, one can have no knowledge of which path the neutron takes which is akin to fully wave-like behaviour. On the other hand, when there is complete absorption of the left beam, $b = 0, \beta = 0$ and $P = 1$. In this case one knows with certainty that the neutron must have taken the right path, and this is particle-like behaviour. The claim is that this therefore provides a generalized quantitative expression for wave-particle complementarity, and one can pass from one extreme to the other by varying the single parameter $\beta$: ``The more clearly we wish to observe the wave nature of light, the more information we must give up about its particle properties'' [2c].

The concept of an entity that is neither fully a wave nor fully a particle is essentially non-classical, and hence such an interpretation is not encompassed by Bohr's wave-particle complementarity interpretation in which the use of incompatible `classical concepts' in a complementary way is at the core, as we have seen. The question is whether Bohr's interpretation is consistent with these welcher-weg experiments. The answer is `yes' beacuse once absorption occurs, the ensemble of neutrons is split into two incoherent parts, one that is absorbed and the rest. The ensemble of unabsorbed neutrons shows completely wave-like behaviour since their paths through the interferometer cannot be determined in principle, whereas the ensemble of absorbed neutrons whose paths can be so determined behaves fully like particles. 

Furthermore, as we have seen, with only one of the two slits fully blocked, what one actually observes is a single-slit diffraction pattern through the other slit which is open, a feature that is not reflected by the relation $P = 1$ which is supposed to hold in this case. It is important therefore to note that the relation  
(\ref{wp}) is not a consequence of quantum mechanics but is an interpretation that is added to it.

\subsection{Delayed-choice experiments}

\begin{figure}[ht]
\begin{picture}(100,100)(-100,0) \put
(-49,1){\line(1,0){175}}\put(1,1){\line(0,1){125}} \put
(1,125){\line(1,0){125}}\put (126,1){\line(0,1){125}}
\put(-15,110){\line(1,1){30}}\put(111,110){\line(1,1){30}}\put(114,110){\line(1,1){30}}
\put(112,-15){\line(1,1){30}}\put(-12,-13){\line(1,1){30}}\put(-9,-13){\line(1,1){30}}
\put(-60,1){S}
\put(-20,-25){BS1}\put(95,100){BS2}
\put(-33,110){$M$}\put(110,-25){$M$}\put(165,121){$D_b$}\put
(125,125){\line(1,0){37}} \put
(120,160){$D_d$}\put(-30,-3){\vector(2,0){10}}
\put(30,-3){\vector(2,0){10}}\put(-3,50){\vector(0,2){10}}
\put(30,122){\vector(2,0){10}}\put(130,50){\vector(0,2){10}}
\put(140,122){\vector(2,0){10}} \put(-40,-50){Fig. 2 A balanced Mach-Zehnder
interferometer}\end{picture}\vspace{1in}\end{figure}

It is important to bear in mind that when unobserved, particles cannot be said to have precise trajectories in space-time, at least in the usual 
Copenhagen interpretation of quantum mechanics. As we have seen, this is a consequence of Heisenberg's uncertainty relation which implies that an atomic object cannot be said to possess simultaneously sharp values of position and momentum. Hence, it is strictly speaking incorrect to talk about `tracing the path' of an atomic object in transit. The incongruity of using the classical concept of a particle to describe an atomic object in transit in quantum mechanics becomes even more apparent in delayed-choice experiments \cite{wheeler}. A typical example is a balanced Mach-Zehnder interferometer set up in which one detector $D_d$ is dark (does not record any photons) and the other detector $D_b$ is bright (records all the photons) (Fig. 2). This is because of destructive interference along the path to the dark detector and constructive interference along the path to the bright detector. If one talks in terms of photons travelling along routes, one has to admit that each photon travels via both routes and interferes with itself, which is unreasonable. If the second beam-splitter is removed, both counters register counts with equal probability. In this case each photon can be said to travel only one route. However, one can decide whether to insert the second beam-splitter in place or to take it out only at the last pecosecond after the photon has already accomplished its travel. In Wheeler's words, 
\begin{quote}
We, now, by moving the mirror $^3$ in or out have an unavoidable effect on what we have a right to say about the {\em already} past history of that photon. 
\end{quote}
This is weird and unaccetable, to say the least. As we will see in the last section, there is no past history of an atomic object in quantum mechanics.

\subsection{Quantum Erasures}
Another example is the {\it quantum erasure} \cite{chiao}. Let me again quote from Gribbin's clear and popular exposition of the double-slit quantum erasure \cite{gribbin}:
\begin{quote}
In this variation on the Young's slit theme, the experiment is first set up in the usual way, and run to produce interference. Quantum theory says that the reason why interference can occur, even if light is a stream of photons, is that there is no way to find out, even in principle, which photon went through which slit. The ``indeterminacy" allows fringes to appear. 

But then Chiao and his colleagues ran the same experiment with polarising filters in front of each of the two slits. Any photon going one way would become ``labelled" with left-handed circular polarization, while any photon going through the other slit is labelled with right-handed circular polarization. In this version of the experiment, it is possible in principle to tell which slit any particular photon arriving at the second screen went through. Sure enough, the interference pattern vanishes -- even though nobody ever actually looks to see which photon went through which slit. 

Now comes the new trick -- the eraser. A third polarising filter is placed between the two slits and the second screen, to scramble up (or erase) the information about which photon went through which hole. Now, once again, it is impossible to tell which path any particular photon arriving at the second screen took through the experiment. And, sure enough, the interference pattern reappears! 

The strange thing is that interference depends on ``single photons'' going through both slits ``at once'', but undetected. So how does a single photon arriving at the first screen know how it ought to behave in order to match the presence or absence of the erasing filter on the other side of the slits?
\end{quote}
These two examples clearly show that Einstein's 1905 hypothesis that light is a stream of photons is hard to reconcile with quantum mechanics. But then one might legitimately ask, `What about the photoelectric and Compton effects?' Strange though it may sound, it has been shown that both these effects can also be explained within quantum mechanics by simply using the interaction of classical light with quantized atoms in the detectors \cite{aspect, ghose}. A simple model of such a detector is an atom with a ground state $\vert g\rangle$ and a continuum of excited states $\vert e\rangle$ with a gap $W_T$. Let its interaction Hamiltonian with a classical electromagnetic field be $\vec{E}.\vec{\hat{D}}$ where $\vec{E}$ is the electric field and $\vec{\hat{D}}$ is the electric dipole operator. Taking $\vec{E} = \vec{E_0} \rm{exp} (i\omega t)$, one obtains the transition rate
\be
\frac{d}{d t} {\cal P}_{g \rightarrow e} = \frac{\pi}{2 \hbar}\vert \langle e \vert \vec{\hat{D}}\vert g\rangle \vert^2 E_0^2 \rho (e)\delta (E_e - E_g - \hbar \omega)
\ee
which contains all the features of photoelectricity: the density of states $\rho(e)$ vanishes if $E_e - E_g < W_T$, and the final energy of the system is $E_e = E_g + \hbar \omega$, and hence the kinetic energy of the emitted electron is $\hbar \omega - W_T$. There is therefore no {\it logical necessity} to introduce the concept of the photon to explain these phenomena. 

It is only fairly recently that the truly quantum nature of light is beginning to be observed definitively through the production of very special states of light like single-photon states and squeezed states \cite{loudon}. Such states can be properly described only in quantum optics which has a different theoretical structure.$^4$ Even there the ontology of the quanta (photons) and the field is a controversial issue \cite{kuhlman}.

\section{CONCLUSIONS}
In all these examples of particle interpretation of the double-slit experiment and its variants there is an implicit assumption that ``an exact description of the past path of a particle'' is possible in quantum mechanics. Recall Bohr's statement quoted in Section 2 that ``it is only the circumstance that we are presented with a choice of {\em either} tracing the path of a particle {\em or} observing interference effects, which allows us to escape from the paradoxical necessity of concluding that the behaviour of an electron or a photon should depend on the presence of a slit in the diaphragm through which it could be proved not to pass''. That ``tracing the path of a particle'' is a classical hangover which quantum mechanics does not permit was first made clear in a paper of Einstein, Tolman and Podolsky \cite{etp} in which they showed that ``the principles of quantum mechanics involve an uncertainty in the description of past events, which is analogous to the uncertainty in the prediction of future events''.  They showed this by applying the time-energy uncertainty relation to a simple set up in which a pair of particles is released from a box containing identical particles in thermal agitation in two directions by means of a shutter. The box is accurately weighed before and after the shutter opens, and one particle travels directly to an observer while the other particle travels a much longer path and is reflected by a parabolic mirror to the observer. The observer measures the momentum of the first particle and then observes its time of arrival. Although the measurement of momentum will change the momentum in an unknown manner, nevertheless knowing its past momentum, and hence its velocity and energy, it would appear possible to know the time when the shutter opened, and hence to calculate the velocity and energy of the second particle from the known loss in the energy content of the box when the shutter opened. It should then be possible to predict both the energy and the time of arrival of the second particle, which is paradoxical because of the time-energy uncertainty relation. The paradox is resolved by noting that the past motion of the particle cannot be accurately determined as assumed. No wonder then that despite much effort Einstein was ``unable to achieve the sharp formulation of Bohr's principle of complementarity''.

The plain truth according to quantum mechanics is therefore that only `process 2' holds between the source and the detector wherever that may be placed at whatever time during the course of the experiment, and `process 1' (or its effective equivalent) takes over only at the place and time of final detection, generating only `which place' or particle-like information. Thus, the localized spots (spots on photographic plates or positions of clicking detectors, for example) are the only directly {\em observable} features of atomic objects  -- their in principle {\em unobservable} propagation between the source and the detector (`process 2') can, if one insists on employing classical concepts, be {\em interpreted} as being wave-like. The two processes of Schr\"{o}dinger evolution (wave-like propagation) and localized detection (collapse or its effective equivalent) are {\em sequential} and {\em complementary} in the {\em same} experiment -- they are mutually exclusive but together they give an exhaustive account of all observable phenomena. Thus, one can maintain both the classical wave concept {\em and} the classical particle concept in {\em every} conceivable experiment, but never both of them simultaneously, only {\em sequentially} in time and with the wave only as an {\em inference} from the statistical pattern of the {\em observed} spots. It is only when wave-particle duality and complementarity are stated in this manner that they appear as obvious elements of a consistent interpretation of standard quantum mechanics. 

It is clear from all this that somehow an extraneous intervention on unitary evolution is required for quantum mechanics not to be vacuous, i.e., devoid of observable phenomena, because there are no `closed' atomic phenomena without irreversible detection and amplification \cite{bohr4}, and the latter processes cannot be a unitary quantum mechanical process like `process 2'. Depending on one's choice, this intervention can be taken to be either von Neumann's `process 1' or Penrose's process ${\bf R}$ or the postulated classicality of the measuring instruments {\em a la} Bohr or something else. The necessity of such an intervention, whatever that might be, constitutes the ``central'' and ``only'' mystery of standard quantum mechanics. 

A number of proposals with different ontologies have been put forward to get round this mystery, this shifty split between the observer and the observed, the most well-known among them being the de Broglie-Bohm causal interpretation in terms of hidden variables \cite{bohm}, Feynman's path-integral formalism \cite{feynman2}, the so-called many worlds interpretation \cite{everett} with its implications for quantum cosmology \cite{gellmann} and quantum computing \cite{deutsch}, the consistent histories approach \cite{griffiths} and decoherence \cite{zurek} and Penrose's proposal that gravity may have a role in quantum state reduction \cite{penrose}. It will be going beyond the scope of this paper to dwell on them.

Let me end with the following provocative and tongue-in-cheek quotation from Bell \cite{bell7}:

\begin{quote}
ORDINARY QUANTUM MECHANICS (as far as I know) IS JUST FINE FOR ALL PRACTICAL PURPOSES. ... So it is convenient to have an abbreviation for the last phrase: FOR ALL PRACTICAL PURPOSES = FAPP. ...

Is it not good to know what follows from what, even if it is not really necessary FAPP? Suppose for example that quantum mechanics were found to {\em resist} precise formulation. Suppose that when formulation beyond FAPP is attempted, we find an unmovable finger obstinately pointing outside the subject, to the mind of the observer, to the Hindu scriptures, to God, or even only Gravitation? Would not that be very, very interesting? 
\end{quote}

\section{ACKNOWLEDGEMENT}

I am most grateful to a number of colleagues, particularly Virendra Singh, Ravi Gomatam and N. Mukunda for helpful comments on the first draft of the paper which have led to considerable expansion and modification.

\vspace{0.1in}
\flushleft
{\bf Notes}
\vspace{0.1in}

{1. This is a Cartesian split. According to Descartes `matter' and `mind' are essentially different substances, the defining attribute of matter being  spatial `extension' and that of mind `self-awareness' and `thought'. Mind is therefore non-physical. He famously wrote, {\em cogito ergo sum} (I think, therefore I am). In his Meditations he says: ``... the human mind is shown to be really distinct from the body, and, nevertheless, to be so closely conjoined therewith, as together to form, as it were, a unity'' (John Veitch Translation of 1901).

2. Here the word `subjective' must be understood in the sense of being intersubjectively valid to avoid solipsism.   

3. The word `mirror' is to be understood here to refer to the `half-silvered mirror' or the second beam-splitter.

4. Such states are defined by a useful convention based on the weight function in the Sudarshan diagonal representation. All photon states and quantized radiation field states are quantum in nature -- we separate them for convenience into what we call `classical' and `non-classical'. -- N. Mukunda (private communication)}

\end{document}